\documentstyle[aps,preprint]{revtex}

\begin{document}
\draft
\preprint{SNUTP/97-111, hep-th/9801063}

\title{Bogomolnyi Bound with a Cosmological Constant }
\author{Yoonbai Kim and Kyoungtae Kimm}
\address{Department of Physics, Sung Kyun Kwan University, 
Suwon 440-746, Korea\\
yoonbai$@$cosmos.skku.ac.kr, dragon$@$newton.skku.ac.kr}
\maketitle

\begin{abstract}
Bogomolnyi-type bound is constructed for the topological solitons in O(3) 
nonlinear $\sigma$ model coupled to gravity with a negative cosmological 
constant. Spacetimes made by self-dual solutions form a class of 
G\"{o}del-type universe. In the limit of a spinless massive point particle, 
the obtained stationary metric does not violate the causality and it is 
a new point particle solution different from the known static hyperboloid 
and black hole. We also showed that static Nielsen-Olesen vortices saturate 
Bogomolnyi-type bound only when the cosmological constant vanishes.
\end{abstract}


Bogomolnyi bound for the system with a scalar potential shows a characteristic
of the vacuum point of zero energy (or zero cosmological constant). This 
observation is guaranteed by obtaining a self-dual system by forcing the 
global supersymmetry. In curved spacetime, one may ask an intriguing question 
of whether or not the zero cosmological constant is 
also a requirement for the Bogomolnyi-type bound. In the context of
supergravity (local supersymmetry) this seems not to be a condition~\cite{AT}. 
Another question is about self-dual solitons in the system with 
constraint but without scalar potential, namely topological lumps in 
O(3) nonlinear $\sigma$ model or Skyrmions. In the case of Skyrmions
Bogomolnyi equation admits nontrivial solutions on the curved three
sphere whereas there does not exist such a self-dual solution
in the flat spacetime~\cite{Man}. 
Until now the obtained Bogomolnyi bounds in curved spacetime 
seem to favor the zero cosmological
constants, and the supersymmetric extension of them has also been analyzed in
Ref.~\cite{Wit}. The Abelian Higgs model and the O(3) nonlinear $\sigma$ model
saturate the Bogomolnyi bound with zero cosmological constant~\cite{CG,GOR}. 
There is a soliton solution of which boundary value does not
go to the vacuum point, and this solution constitutes compact two dimensional
spatial manifold, specifically two sphere~\cite{Lin}. 
{}For the spinning case, e.g., Chern-Simons Higgs model, 
the scalar potential of the self-dual system in
curved spacetime asks negative $\phi^{8}$ terms of which coefficient is
proportional to the Newton's constant in addition to $\phi^{6}$ terms for flat
spacetime case~\cite{Val}. 
However, the boundary values for all finite-energy regular
self-dual solitons are obtained in open spacetime 
to have the zero cosmological constant even
under the scalar potential unbounded below~\cite{CCK}.
Here we will investigate Bogomolnyi bound of O(3) nonlinear $\sigma$ model
with a nonvanishing cosmological constant.

In this paper, we will study what kind of spacetime geometry is  allowed
with a nonzero cosmological constant and with a few reasonable matter sources.
The matter distribution of our interest is given by the 
static self-dual solitons
of O(3) nonlinear $\sigma$ model, and thereby the only nonvanishing
component of the energy momentum tensor is energy density. 
This means that the obtained topological lumps saturate  so-called
Bogomolnyi-type bound in a curved spacetime with a nonvanishing vacuum
energy. Though the matter distribution is static and does not carry both
internal and external angular momentum, the corresponding metric is stationary 
due to the nonvanishing vacuum energy. 
In the limit of a point particle, the obtained stationary solution implies the
existence of a new metric different from the known (2+1)-dimensional [(2+1)D] 
anti de-Sitter solutions, i.e., regular hyperboloid 
with a deficit angle~\cite{DJ} 
or Ba\~{n}ados-Teitelboim-Zanelli (BTZ) type black hole~\cite{BTZ}. 
Since the latter two solutions cover all
possible solutions of static metric \cite{DJ,KKK}, all (2+1)D anti-de Sitter
solutions formed by massive point particles are classified by these.
On the other hand, the limit of a constant energy density leads to 
one parameter family of the G\"{o}del-type solutions 
with including closed timelike curves~\cite{God}.

Einstein gravity in 2+1 dimensions is described by the action,
\begin{eqnarray}
S=-\frac{1}{16\pi G} \int d^3x \sqrt{g}~(R+ 2 \Lambda) 
+S_{\rm matter}.
\end{eqnarray}
{}For static matter distributions, the most general form of
metric is stationary:
\begin{eqnarray}
\label{smet}
ds^2=N^2(dt +K_i dx^i)^2-\gamma_{ij} dx^i dx^j ,
\end{eqnarray}
where $N^2$, $K_i$, and $\gamma_{ij}~(i,j=1,2)$ are the functions of spatial 
coordinates and $\gamma_{ij}$ depicts spacelike hypersurface.
{}From now on, we use $\gamma_{ij}$ in raising or lowering all spatial indices.

A matter action of our interest is the O(3) nonlinear $\sigma$ model
which admits static topological solitons and saturate 
Bogomolnyi limit in flat spacetime. Once we assume the 
self-dual solitons which saturate Bogomolnyi-type bound in relativistic
models, a characteristic is the vanishment of $ij$-components of energy
momentum tensor:
\begin{eqnarray}
T^{ij}=0,
\end{eqnarray}
and the superposition of the energy of each unit soliton (here we will call
$\int d^{2}x\sqrt{g}T^{t}_{\;t}$ as the energy for simplicity) for multi-center
static solitons. In addition, if the solitons do not carry angular momentum,
then they have
\begin{eqnarray}
T^{i}_{\; t}=0.
\end{eqnarray}

{}For a matter distribution with nonvanishing energy density $T^t_{\;t}$,
the $tt$-component of Einstein equation is 
\begin{eqnarray}\label{00eq}
-\frac{2}{b}\partial_{z}\partial_{\bar{z}}\ln{}b+\frac{3}{4}N^2 K^2
-\frac{1}{2N}\epsilon^{ij}K_i\nabla_j(N^3 K) -\Lambda=8\pi GT^{t}_{\;t},
\end{eqnarray}
where $\nabla_i$ is 2D covariant derivative,
$K=\epsilon^{ij}\nabla_i K_j$, and we have used conformal gauge
$\gamma_{ij}=b(z,\bar{z})\delta_{ij}$ with $z=x+iy$.
Above all, we solve exactly $it$-components of the Einstein equation 
$(G^{i}_{\;t} =0)$ and obtain 
\begin{eqnarray}
\label{0ieq}
N^3K=\kappa,
\end{eqnarray}
where $\kappa$ is an arbitrary integration constant.
After some calculation, the traceless and trace part of
$ij$-components of the Einstein equation become as follows
\begin{eqnarray}
\label{stless}
&&\frac{1}{b}\partial_{\bar z} N=V(z), \\
&&\frac{1}{b}\partial_{\bar z} N\Big( 4\partial_z \partial_{\bar z}\ln b
+2\Lambda b -\frac{3\kappa^2}{2N^{4}}b\Big)=0,
\label{strace} \\
&&(\mbox{\rm or equivalently}
\quad\frac{4}{b}\partial_{z}\partial_{\bar{z}}N
+2\Lambda N+\frac{\kappa^2}{2N^{3}}=0),
\label{Neqst}
\end{eqnarray}
where $V(z)$ is an arbitrary holomorphic function.
Equation~(\ref{strace}) tells us that solutions should satisfy either 
$4\partial_z \partial_{\bar z}\ln b+2\Lambda b -3\kappa^2b/2N^{4}=0$
or $V(z)=\partial_{z}N/b=0$.
The consistency between Eqs.~(\ref{00eq}) and (\ref{strace}) provides us a
constraint that $V(z)$ should vanish at any point where $T^{t}_{\;t}$ does not
vanish. Obviously the easiest way is to look for static solutions with
vanishing $\kappa$ and off-diagonal components of the metric functions vanish,
i.e., $K_i=0$ up to a gauge degree from Eq.~(\ref{0ieq}). 
Since all nonlinear terms in Eqs.~(\ref{00eq}) and (\ref{Neqst}) disappear in
the $\kappa=0$ case, an approach in this direction with a
cosmological constant was firstly done in Ref.~\cite{DJ} and the obtained 
solutions has been interpreted as a two sphere in de Sitter space, and
a hyperboloid in anti-de Sitter space. Schwarzschild-type black hole structure
was recently reported when the cosmological constant is negative~\cite{BTZ},
and those solutions are actually half of anti-de Sitter solutions obtained
in Ref.~\cite{DJ} but were never tried to be interpreted 
in that paper \cite{KKK}.
At first glance, the above solutions compatible with static metric may look
to be the only natural solutions for static spinless sources. On the other
hand, there is no {\em a priori} reason to set 
the integration constant $\kappa$ in Eq.~(\ref{0ieq}) to be zero. 

In this paper, let us examine another case, 
i.e., $V(z)=0$, which must be applicable to a construction of 
Bogomolnyi bound with a cosmological constant. 
Since $b$ is a nonvanishing real function
and $N$ is also a real metric function, equation~(\ref{stless}) forces
$N^2=|\Lambda|^{-1}$ up to rescaling of time.
{}From Eqs.~(\ref{0ieq}) and (\ref{strace}), we fix 
the integration constant $\kappa$,
\begin{eqnarray}
\label{const2}
\kappa^2 =-\frac{4}{\Lambda}\quad(\Lambda \le 0), 
\end{eqnarray}
and then the metric functions $K_{i}$ are determined by solving an equation:
\begin{eqnarray}\label{offd}
K=2{|\Lambda|}.
\end{eqnarray}
Note that when $N=const$ in anti-de Sitter spacetime, the metric $K_{i}$ should 
not vanish despite of the static matter distribution. 
Substituting $N^2=|\Lambda|^{-1}$ and Eq.~(\ref{const2}) into Eq.~(\ref{00eq}),
we have 
\begin{eqnarray}\label{muls}
\partial_{z}\partial_{\bar{z}}\ln b -2|\Lambda| b =-4\pi G b T^{t}_{\;t}.
\end{eqnarray}

In the limit of negligible core size of such localized self-dual solitons, 
dynamics in curved spacetime can be replaced 
by that of spinless point particles at rest. The matter action  
of the point particles is 
\begin{eqnarray}
S_{\rm matter}=\sum_{a} m_{a}\int^{\infty}_{-\infty}ds\sqrt{g_{\mu\nu}
\frac{dx^{\mu}_{a}}{ds}\frac{dx^{\nu}_{a}}{ds}}.
\end{eqnarray}
Though we can have multi-particle solutions of 
Eq.~(\ref{muls}), we now restrict our interest to rotationally
symmetric solutions of Eq.~(\ref{muls}),
\begin{eqnarray}
b&=&\frac{\alpha^2}{|\Lambda|r^2[(r/r_0)^\alpha-(r_0/r)^\alpha]^2},
\end{eqnarray}
where the constant $\alpha$ is determined to give the correct right-hand 
side of (\ref{muls}).  Off-diagonal components of the metric are also 
obtained from Eq.~(\ref{offd}), and we have a stationary metric for 
the static massive but spinless particles sitting in curved spacetime 
with a negative cosmological constant:
\begin{eqnarray}\label{ptmet}
ds^{2}=\frac{1}{|\Lambda|}\left(dt- 
\frac{\alpha(r/r_0)^\alpha}{(r/r_0)^\alpha-(r_0/r)^\alpha} d\theta\right)^2
-\frac{\alpha^2(dr^2+r^2d\theta^2)}{
|\Lambda|r^2[(r/r_0)^\alpha-(r_0/r)^\alpha]^2} ,
\end{eqnarray}
where the valid range of radial coordinate is $0\le r < r_0$.
If we make a coordinate transformation which uses the proper spatial distance
as a radial coordinate, i.e.,
$\tanh\chi=(r/r_{0})^{\alpha}$ $(0\leq \chi\leq\infty$, $\alpha=1-4Gm)$, 
and $\vartheta=(1-4Gm)\theta$, the metric (\ref{ptmet}) is rewritten by
\begin{eqnarray}\label{reptm}
|\Lambda|ds^{2}=\left[dt+\sinh^{2}\chi{}d\vartheta\right]^2
-d\chi^{2}-\frac{1}{4}\sinh^{2}(2\chi)d\vartheta^{2}.
\end{eqnarray}
The spatial part of the above metric is nothing but that of a hyperboloid
embedded in 3-dimensional Minkowski space, which has a deficit 
angle proportional to the point particle mass located at the origin.
In fact, this metric is an analog of the
3-dimensional G\"{o}del type cosmological model~\cite{God}. 
However it should be emphasized that the spacetime of Eq.~(\ref{reptm})
does not violate causality, i.e., there is no closed time-like curves.
If we calculate curvature tensor in the orthonormal frame given by
$e^{\hat{t}}=[dt+\sinh^2\chi]/\sqrt{|\Lambda|}$, 
$e^{\hat{\chi}}=d\chi/\sqrt{|\Lambda|}$, 
$e^{\hat{\vartheta}}=\sinh(2\chi)d\vartheta/2\sqrt{|\Lambda|}$,
then we have 
$R_{\hat{\mu}}{}^{\hat{\nu}}={\rm diag}(2,2,2)$.
This clearly shows that the spacetime is homogeneous and 
has constant curvature except for the origin.

Though the structure of the spatial manifold is the same as that obtained in 
Ref.~\cite{DJ}, the existence of the $t$-$\vartheta$ cross term implies a
difference in dynamics of both classical and quantum test particles. For
a classical test particle, let us look for geodesic equation 
\begin{eqnarray}
\varepsilon=-\dot{\chi}^{2}+\gamma^{2}-\tanh^{2}\chi\left(\gamma
-\frac{\ell}{\sinh^{2}\chi}\right)^{2},
\end{eqnarray}
where $\varepsilon=g_{\mu\nu}\dot{x}^\mu\dot{x}^\nu$ is 
the mass of the test particle which can be rescaled to zero or 
one, and $\gamma$ and $\ell$ are two constants of 
motion for two killing vectors $\partial/\partial t$, 
$\partial/\partial \vartheta$ respectively 
(all quantities were rescaled with respect to $\sqrt{|\Lambda|}$):
\begin{eqnarray}
\dot{t}=\frac{\gamma+\ell}{\cosh^{2}\chi}
\quad\mbox{and}\quad
\dot{\vartheta}=\frac{1}{\cosh^{2}\chi}
\left(\gamma-\frac{\ell}{\sinh^{2}\chi}\right).
\end{eqnarray}
The allowed motions of the massive test particle are circular orbit or 
the bound orbit between the perihelion and aphelion, however those of the
massless test particle are unbounded.
The orbits are  explicitly given by
\begin{eqnarray}
\cos(\vartheta-\vartheta_0)=
2\frac{(\gamma+2\ell)\sinh^2\chi+\ell}
{\sqrt{(\gamma+2\ell)^2-\varepsilon}\sinh(2\chi)}.
\end{eqnarray}
When it does not carry the angular momentum $\ell$,  
the massless test particle injected from 
$\chi=\infty$ with zero impact parameter is
deflected by $90$ degree irrespective of its mass $\varepsilon$ and energy
$\gamma$ when it arrives at $\chi=0$. 
On the other hand, the $\ell\ne 0$ massless test particle keeps straight when
it reaches the perihelion at 
$\chi_{\rm peri}=\sinh^{-1}(\ell/\gamma\sqrt{1+2\ell/\gamma})$. 
If we write down the energy-momentum tensor with inclusion of the vacuum
energy term such as
\begin{eqnarray}
\tilde{T}^{\mu\nu}=T^{\mu\nu}+\frac{\Lambda}{8\pi G} g^{\mu\nu}, 
\end{eqnarray}
we easily notice that the negative cosmological constant term plays
effectively a role of angular momentum:
\begin{eqnarray}
J=\int d^2x \epsilon_{ij} x^i \tilde{T}^{tj}
=\frac{\Lambda}{8\pi G}\int dx^2\epsilon_{ij} x^i g^{tj}\ne 0.
\end{eqnarray}
In this sense the negative cosmological constant provides a rotating frame 
to the test particle and its role resembles that of background 
electromagnetic field to a charged particle.
The quantum physics in the background gravity of which the background is 
provided by (\ref{reptm}) is expected to raise a question of  
whether the energy is quantized or the angular momentum similar 
to the case of spinning strings
given by the 3-dimensional ``Kerr solution''~\cite{Maz}.
However, technical difficulty in the separation of variables hinders detailed
analysis.

The opposite limit is to let the  size of soliton become infinite, 
and then the energy-momentum tensor is expressed by
\begin{eqnarray}
T^{\mu\nu}=\rho V^\mu V^\nu ,
\end{eqnarray}
where $\rho$ is a constant energy density and $V^{\mu}=\delta^\mu_{t}$.
The $it$- and $ij$-components of Einstein equation are the same as those of
the points particles except for the normalization $N^2=|\beta\Lambda|^{-1}$, 
and remaining temporal component of the Einstein equation  
is also reduced to a Liouville equation
\begin{eqnarray}
\partial_{z}\partial_{\bar{z}} \ln b =2\beta|\Lambda| b, 
\end{eqnarray}
where $\beta\equiv 1-2\pi G\rho/|\Lambda|$.
Rotationally symmetric
solutions are classified by the value of $\beta$: When $\beta=0$, the line
element is flat such as 
\begin{eqnarray}
ds^2=\frac{1}{|\Lambda|}(dt-r^{2}d\theta)^{2}-\frac{1}{|\Lambda|}
(dr^{2}+r^{2}d\theta^{2}).
\end{eqnarray}
When $\beta\ne 0$, it becomes
\begin{eqnarray}\label{metG}
ds^2=\frac{1}{|\beta\Lambda|}\left(dt-
\frac{r/r_0d\theta}{\sqrt{|\beta|}(r/r_0\mp r_0/r)}\right)^2
-\frac{1}{|\beta\Lambda|}
\frac{dr^2+r^2d\theta^2}{r^2(r/r_0\mp r_0/r)^2} .
\end{eqnarray}
Once again let us do a coordinate transformation. For $\beta>0$,
$r/r_{0}=\tanh\chi$ and then we have
\begin{eqnarray}\label{Godel}
|\beta\Lambda|ds^2&=&\left(dt
+\frac{1}{\sqrt\beta}\sinh^2\chi
d\theta\right)^2
-d\chi^2 -\frac{1}{4}\sinh^2(2\chi)d\theta^2.
\end{eqnarray}
This metric describes one parameter family of G\"{o}del-type universe in
3-dimensions. Unlike the spacetime of the point particle it has no 
deficit angle.
For $\beta<0$, $r/r_{0}=\tan\chi$ and the metric (\ref{metG}) is
transformed to
\begin{eqnarray}\label{cptsp}
|\beta\Lambda|ds^2&=&
\left(dt+\frac{1}{\sqrt{|\beta|}}\sin^{2}\chi
d\theta\right)^2-d\chi^2 -\frac{1}{4}\sin^{2}(2\chi) d\theta^2.
\end{eqnarray}
Note that the above metric in Eq.~(\ref{Godel}) depicts the original 
G\"{o}del universe when $\beta=1/2$ \cite{God}. 
Introducing an  orthonormal frame
$e^{\hat{t}}=
[dt+(1/\sqrt{\beta})\sinh^2\chi{}d\theta]/\sqrt{|\beta\Lambda|}$,
$e^{\hat{\chi}}=d\chi/\sqrt{|\beta\Lambda|}$,
$e^{\hat{\vartheta}}=\sinh(2\chi)d\theta/2\sqrt{|\beta\Lambda|}$, we have
constant curvature
$R_{\hat\mu}{}^{\hat\nu}={\rm diag}(2,4-2/\beta,4-2/\beta)$.
Although this metric describes a
homogeneous universe, it contains  closed time-like curves for the range,
$\tanh\chi >\sqrt{\beta}$.
In addition to two Killing vectors, $\partial/\partial{}t$,
$\partial/\partial\theta$ corresponding to time translation and rotational
invariance, we can find two more Killing vectors:
\begin{eqnarray}
X_{(1)}&=&
\frac{1}{\sqrt{\beta}}\tanh\chi\sin\theta\frac{\partial}{\partial t}
       -\cos\theta\frac{\partial}{\partial \chi}
       +2\coth(2\chi)\sin\theta\frac{\partial}{\partial\theta},\\
X_{(2)}&=&
\frac{1}{\sqrt{\beta}}\tanh\chi\cos\theta\frac{\partial}{\partial t}
       +\sin\theta\frac{\partial}{\partial \chi}
       +2\coth(2\chi)\cos\theta\frac{\partial}{\partial\theta}.
\end{eqnarray}
These four Killing vectors satisfy the following so(2,1) algebra:
\begin{eqnarray}
\left[ X_{(0)},X_{(i)}\right]&=& 0\quad (i=1,2,3), \nonumber \\
\left[ X_{(3)},X_{(1)}\right]&=&2X_{(2)},\quad
\left[ X_{(3)},X_{(2)}\right]= -2X_{(1)},\\
\left[ X_{(1)},X_{(2)}\right]&=&-2X_{(3)}\nonumber ,
\end{eqnarray}
where $X_{(0)}=\partial/\partial{}t$,
$X_{(3)}=(1/\sqrt{\beta})\partial/\partial{}t +2\partial/\partial\theta$.
When $\beta=1$ this set of Killing vectors coincides with that of the metric
(\ref{reptm}).

Now we consider O(3) nonlinear $\sigma$ model 
described by the matter action
\begin{eqnarray}
\label{saction}
S_{\sigma}=\int d^3x\sqrt{g}
\frac{\mu^2}{2}\partial_\mu \phi^a \partial^\mu \phi^a 
\quad(\phi^a\phi^a=1,\quad{}a=1,2,3), 
\end{eqnarray}
where we have inserted mass scale $\mu$. Similarly, $it$-components
of the energy-momentum tensor vanish for any static objects in this model. 
Now we concentrate on the static regular self-dual solitons,
so to speak the extended objects $\phi^a$ satisfying the self-dual
equation
\begin{eqnarray}
\label{self_s} 
\partial_i \phi^a \pm \epsilon_i{}^j \varepsilon^{abc}\phi^b\partial_j \phi^c=0.
\end{eqnarray}
A characteristic of self-dual solitons is the vanishment of 
$ij$-components of the energy-momentum tensors:
\begin{eqnarray}
T^{ij}=
\frac{\mu^2}{4}\Big\{
(\partial^i \phi^a \pm \epsilon^i{}_k\epsilon^{abc}\phi^b\partial^k \phi^c )
(\partial^j \phi^a \mp \epsilon^j{}_l\epsilon^{abc}\phi^b\partial^l \phi^c )
+(i\leftrightarrow j)\Big\}=0.
\end{eqnarray}
A requirement is the reproduction of Euler-Lagrange equation from the 
Bogomolnyi equation~(\ref{self_s}), and then we get a constraint
\begin{eqnarray}
\label{const1}
N^{-1}\left[ V(z)\partial_z 
+\bar{V}({\bar z})\partial_{\bar z}\right]\phi^a=0.
\end{eqnarray}
Suppose that $V(z)$ does not vanish. Rewriting 
the above equation (\ref{const1}) by introducing a real variable 
$\zeta_1=(1/2)[\int^z dw/V(w)+\int^{\bar z} d\bar w /\bar{V}(\bar w)]$
such as $N^{-1}\partial_{\zeta_{1}}\phi^a=0$,
one may conclude that any $\phi^a$-configuration independent of $\zeta_1$
cannot describe nontrivial soliton solutions. 
Therefore $V(z)$ should be zero everywhere, the whole procedure is repeated as
has been done in the previous cases.
If we substitute $N^2=|\Lambda|^{-1}$ and Eq.~(\ref{const2}) into 
Eq.~(\ref{00eq}), the Einstein equation is expressed by
\begin{eqnarray}
\label{s00eq2}
\partial_{z}\partial_{\bar{z}}\ln b -2|\Lambda|b=-\pi G\mu^2 b\epsilon^{ij}
\varepsilon^{abc} \phi^a \partial_i \phi^b\partial_j \phi^c.
\end{eqnarray}
Note that the remaining equations are the self-dual equation
({\ref{self_s}) and the $tt$-component of Einstein equation (\ref{s00eq2}).
Thus we now obtain Bogomolnyi-type bound of the O(3) nonlinear
$\sigma$ model when the cosmological constant is nonpositive.
For convenience we use stereographic projection of 
$\phi^a$ field onto complex $u$-plane:
\begin{eqnarray}\label{stere}
u= \frac{\phi^1 +i \phi^2}{1+\phi^3}.
\end{eqnarray}
Inserting Eq.~(\ref{stere}) into
Eq.~(\ref{s00eq2}), we have
\begin{eqnarray}\label{o3ein}
\partial_{z}\partial_{\bar{z}}\ln b -2|\Lambda|b
=-8\pi G \mu^2 \partial_{z}\partial_{\bar{z}} \ln (1+\bar{u}u).
\end{eqnarray}
Self-dual solutions of Eq.~(\ref{self_s}) are given by (anti-)holomorphic 
function and,
specifically, the most general solution corresponding topological
charge $n$ is 
\begin{eqnarray}
u=c \prod_{i=1}^{n}\frac{z-a_i}{z-b_i},
\end{eqnarray}
where $a_i$ and $b_i$ are complex numbers. Let us consider the soliton solution 
of unit topological charge, i.e., $u=z_0/z$. 
Only when $2\pi G\mu^2=1$, an exact solution of Eq.~(\ref{o3ein}) is obtained,
\begin{eqnarray}
b=\frac{1}{|\Lambda|r^{2}_{0}[1+(r/r_{0})^{2}]^2} .
\end{eqnarray}
By solving Eq.~(\ref{offd}) again, the line element becomes
\begin{eqnarray}
ds^2=
\frac{1}{|\Lambda|}\left(dt+\frac{(r/r_0)^2d\theta}{1+(r/r_{0})^{2}}
\right)^2 
-\frac{dr^2 +r^2d\theta^2}{|\Lambda|r_{0}^{2}[1+(r/r_{0})^{2}]^2}.
\end{eqnarray}
Transforming the radial coordinate to the proper spatial radius $\chi$, i.e., 
$r/r_{0}=\tan\chi$ $(0\le\chi\le \pi/2)$, we obtain
\begin{eqnarray} \label{cirm}
|\Lambda|ds^{2}
&=&\left(dt +{2}\sin^2\chi{}d\theta\right)^2
-d\chi^2 -\frac{1}{4}\sin^2(2\chi)d\theta^2 .
\end{eqnarray}
This metric corresponds to the $\beta=-1/4$ case of the constant energy
density in Eq.~(\ref{cptsp}) and $\gamma_{ij}$ part of 
Eq.~(\ref{cirm}) depicts a perfect two sphere.
The existence of the closed timelike curves of this metric can be easily read
from the $g_{\theta\theta}$. Indeed, for the range
$\pi/4<\chi<\pi/2$, the circles $t=\chi=const$ are closed
timelike curves. 

When $2\pi G \mu^2<1$, the asymptotic form of the metric  function
$b(r)$ behaves $1/r^4$ and it implies that its $\gamma_{ij}$ part also
describes a 2D manifold which is topologically equivalent to $S^2$.
The causality violation in the metric is manifest due to 
the relation in Eq.~(\ref{offd}).

Though we obtained static non-self-dual solitons of static metric including
charged BTZ black holes~\cite{KM}, and static self-dual solitons of a
stationary metric in this paper, 
we do not know whether or not there is static non-self-dual
solitons of a stationary metric in anti-de Sitter spacetime, yet.

Another intriguing soliton to saturate the Bogomolnyi-type bound is
the Nielsen-Olesen vortex in Abelian Higgs model described by the action 
\begin{eqnarray}
S_{AH}=\int d^3x \sqrt{g}\bigg\{
-\frac{1}{16\pi G} (R+2\Lambda)
-\frac{1}{4}F_{\mu\nu}F^{\mu\nu}
+(D_\mu \phi)^* (D^\mu\phi) -V(\phi)
\bigg\},
\end{eqnarray}
where $D_\mu \phi =(\partial_\mu-ie A_\mu)\phi$. 
In curved spacetime, the Bogomolnyi bound has also been achieved when the 
cosmological constant
is zero \cite{CG}. Similar to the self-dual solitons in the O(3) nonlinear
$\sigma$ model, these static self-dual vortices are spinless $(T^{i}_{\;t}=0)$
and stressless ($T^{ij}=0$) if they saturate the Bogomolnyi bound. The whole
story for the Einstein equations is the same as the previous procedure for the
self-dual solitons in the O(3) nonlinear $\sigma$ model. However, the last 
condition remains that
the self-dual equations should reproduce Euler-Lagrange equations.
If they are static electrically-neutral $(F_{it}=0)$ vortices with quantized
magnetic flux, Gauss' law gives a constraint
\begin{eqnarray}
N^{2}K\epsilon^{ij}F_{ij}=0.
\end{eqnarray}
Since vortices carry nonzero magnetic field, we have $N^{2}K=0$ which cannot
be consistent with Eqs.~(\ref{0ieq}) and (\ref{const2}) for the case of 
nonvanishing
cosmological constant. Therefore, we lead to the conclusion that the
static magnetically-charged vortices cannot saturate the Bogomolnyi bound in
both de Sitter and anti-de Sitter spacetime.

\vskip 5mm
This work was supported by the Ministry of Education (BSRI/97-2418), the KOSEF
(Grant No. 95-0702-04-01-3), and Faculty Research Fund, Sung Kyun Kwan
University, 1997.

\end{document}